\pdfoutput=1
\documentclass[aps,prd,amsmath,floats,floatfix, twocolumn,
superscriptaddress,nofootinbib,showpacs,longbibliography]{revtex4-1}

\usepackage[T1]{fontenc}
\usepackage[utf8]{inputenc}
\usepackage{lmodern}

\usepackage{verbatim}

\usepackage[dvipsnames, usenames]{xcolor}
\definecolor{linkcolor}{rgb}{0.0,0.3,0.5}
\usepackage[hypertexnames=false, unicode, colorlinks=true, linkcolor=linkcolor,
citecolor=linkcolor, filecolor=linkcolor,urlcolor=linkcolor,
pdfusetitle]{hyperref}

\usepackage[all]{hypcap}
\usepackage{graphicx}
\usepackage{xspace}
\usepackage{amssymb}
\usepackage[normalem]{ulem}
\usepackage{bm}

\usepackage{microtype}

\graphicspath{%
  {figs/}%
}

\DeclareMathAlphabet{\mathpzc}{OT1}{pzc}{m}{it}

\newcommand{\roughly}{\mathchar"5218\relax\,}
\newcommand{\into}{\!\times\!\relax}

\newcommand{\h}{\mathpzc{h}}

\newcommand{\hlm}{\mathpzc{h}_{\ell m}}

\newcommand{\chieffCoorb}{\chi^{\mathrm{coorb}}_{\mathrm{eff}}}

\newcommand{\bchi}{\bm{\chi}}
\newcommand{\bchiCoorb}{\bm{\chi}^{\mathrm{coorb}}}
\newcommand{\chiCoorb}{\chi^{\mathrm{coorb}}}
\newcommand{\chihatCoorb}{\hat{\chi}^{\mathrm{coorb}}}

\newcommand{\bLambda}{\bm{\Lambda}}

\newcommand{\tmHundred}{t\!=\!-100M}

\newcommand{\ts}{\mathcal{TS}}
\newcommand{\ps}{\mathcal{PS}}

\begin{document}

\title{Gravitational wave peak luminosity model for precessing binary black
holes}

\newcommand\caltech{\affiliation{TAPIR 350-17, California Institute of
Technology, 1200 E California Boulevard, Pasadena, CA 91125, USA}}

\newcommand\cornellVarma{\affiliation{Department of Physics, and Cornell
    Center for Astrophysics and Planetary Science, Cornell University, Ithaca,
    New York 14853, USA}}

\author{Afura Taylor}
\email{afura.taylor@gmail.com}
\caltech

\author{Vijay Varma}
\email{vvarma@cornell.edu}
\thanks{Klarman fellow}
\cornellVarma
\caltech

\hypersetup{pdfauthor={Taylor et al.}}

\date{\today}

\begin{abstract}
When two black holes merge, a tremendous amount of energy is released in the
form of gravitational radiation in a short span of time, making such events
among the most luminous phenomenon in the universe. Models that predict the
peak luminosity of black hole mergers are of interest to the gravitational wave
community, with potential applications in tests of general relativity. We
present a surrogate model for the peak luminosity that is directly trained on
numerical relativity simulations of precessing binary black holes. Using
Gaussian process regression, we interpolate the peak luminosity in the
7-dimensional parameter space of precessing binaries with mass ratios $q\leq4$,
and spin magnitudes $\chi_1,\chi_2\leq0.8$. We demonstrate that our errors in
estimating the peak luminosity are lower than those of existing fitting
formulae by about an order of magnitude. In addition, we construct a model for
the peak luminosity of aligned-spin binaries with mass ratios $q\leq8$, and
spin magnitudes $|\chi_{1z}|,|\chi_{2z}|\leq0.8$. We apply our precessing model
to infer the peak luminosity of the GW event GW190521, and find the results to
be consistent with previous predictions.
\end{abstract}

\maketitle

\section{Introduction}
\label{sec:introduction}

As the gravitational wave (GW) detectors LIGO~\cite{TheLIGOScientific:2014jea}
and Virgo~\cite{TheVirgo:2014hva} approach their design sensitivity, GW
detections are becoming routine~\cite{LIGOScientific:2018mvr, Abbott:2020uma,
LIGOScientific:2020stg, Abbott:2020khf, Abbott:2020tfl}. Binary black hole
(BBH) mergers are the most abundant source for these detectors. Such mergers
provide a unique laboratory for studying black hole (BH) astrophysics as well
as for testing general relativity. At the time of merger, the BHs are moving at
about half the speed of light and the spacetime is highly dynamical. As a
result, for a brief moment, BBH mergers are among the most luminous events in
the universe. For example, the recently announced GW event
GW190521~\cite{Abbott:2020tfl} radiated $\sim7.6M_{\odot}$ of energy in GWs in
a fraction of a second, reaching a peak luminosity of $\sim 208 \, M_{\odot} \,
c^2/s = 3.7\times10^{56}$ erg/s~\cite{Abbott:2020mjq}.

The above estimate is obtained by applying peak luminosity
models~\cite{Keitel:2016krm, Healy:2016lce} based on numerical relativity (NR)
simulations to the measured masses and spins of the component BHs.  Apart from
predicting the peak luminosity of GW events, such models can be used to
understand the impact of supermassive BH mergers on circumbinary accretion
disks~\cite{Schnittman:2013qxa} and possible electromagnetic
counterparts~\cite{Kocsis:2008aa, Li:2012dta}. In addition, one can test
general relativity by independently estimating the peak luminosity through a
theory-independent signal reconstruction~\cite{Cornish:2014kda,
Millhouse:2018dgi} and comparing with the prediction from NR. A similar test
was performed for the peak frequency in Ref.~\cite{Carullo:2018gah}. As
detector sensitivity improves, these applications will need accurate models
that capture the full physics of the NR simulations.

NR simulations are essential to model the BH merger process and the resulting
GW peak luminosity. However, these are prohibitively expensive for most GW data
analysis applications. As a result, various phenomenological fits have been
developed for the peak luminosity~\cite{Keitel:2016krm, Healy:2016lce,
JimenezForteza2016, Baker:2008mj}; starting with an ansatze, these models
calibrate any free coefficients to NR simulations. However, all of these models
are restricted to aligned-spin systems, where the BH spins are aligned to the
orbital angular momentum direction ($\hat{\bm{L}}$). For generic binaries,
however, the spins can be titled w.r.t. $\hat{\bm{L}}$. For these systems, the
spins interact with the orbit (and each other), leading to precession of the
orbital plane and the spins~\cite{Apostolatos:1994pre}. Precession causes
modulations in the GW signal and as a result the peak luminosity is affected.

In this paper, we present a Gaussian process regression (GPR) based NR
surrogate model for the peak luminosity of generically precessing BBHs. NR
surrogate models directly interpolate between NR simulations rather than assume
an ansatze about the underlying phenomenology. These methods have been
successfully used to model the GW signal~\cite{Varma:2019csw, Blackman:2017pcm,
Blackman:2017dfb} as well as the remnant BH properties~\cite{Varma:2018aht,
Varma:2019csw, Varma:2020nbm} of precessing BBHs. Through cross-validation
studies, these models have been shown to approach the accuracy level of the NR
simulations themselves.

In particular, we present two models:
\begin{enumerate}
\item \texttt{NRSur7dq4Remnant}: a 7-dimensional precessing model trained
    against systems with mass ratios $q\leq4$, ~\footnote{We use the convention
    $q=m_1/m_2 \geq 1$, where $m_1$ ($m_2$) is the mass of the heavier
    (lighter) BH.} dimensionless spin magnitudes $\chi_{1},\chi_{2} \leq0.8$,
    and generic spins directions.
\item \texttt{NRSur3dq8Remnant}: a 3-dimensional aligned-spin model trained
    against systems with mass ratios up to $q\leq8$ and aligned-spins
    $|\chi_{1z}|,|\chi_{2z}|\leq0.8$.
\end{enumerate}
We use the same names, respectively, as the precessing remnant model of
Ref.~\cite{Varma:2019csw} and the aligned-spin remnant model of
Ref.~\cite{Varma:2018aht}, as we make the models available in the same
interface through the publicly available Python module
\texttt{surfinBH}~\cite{surfinBH}. Even though peak luminosity is not
technically a property of the remnant BH, we expect that using the same
interface will make using the models easier for our users.

The rest of the paper is as follows. We describe our fitting procedure in
Sec.~\ref{sec:methods}. In Sec.~\ref{sec:errors}, we compare the models against
NR simulations to assess their accuracy. In Sec.~\ref{sec:GW190521_peak_lum}, we
apply our precessing model to predict the peak luminosity of GW190521.  We end
with some concluding remarks in Sec.~\ref{sec:conclusion}.

\section{Modeling methods}
\label{sec:methods}

The GW luminosity is defined as~\cite{Baker:2008mj} :
\begin{gather}
\label{eq:luminosity}
    \mathcal{L}(t) = \frac{1}{16\pi} \sum_{\ell, m}
    \left| \lim\limits_{r \to \infty} \left( r \, \dot{\h}_{\ell m}
    \right)\right|^2 ,
\end{gather}
where the dot represents a time derivative, the $|~|$ represents the absolute
value, and $\hlm$ represents the complex spin $\!=\!\!-2$ weighted spherical
harmonic mode with indices $(\ell,m)$. We use the time derivative of $r \hlm$
extrapolated to future null infinity~\cite{Boyle:2009vi} in the place of
$\lim\limits_{r \to \infty} \left( r \, \dot{\h}_{\ell m} \right)$. The
extrapolated strain data is obtained from NR simulations performed with the
Spectral Einstein Code (SpEC)~\cite{SpECwebsite} code, available through the
Simulating eXtreme Spacetimes (SXS)~\cite{SXSWebsite}
Catalog~\cite{Boyle:2019kee, SXSCatalog}. The strain data is first
interpolated onto a uniform time array (with step size $0.1M$, where $M$ is the
total mass) using cubic splines. Then we use a fourth-order finite-difference
derivative to get the time derivative of the strain.

We determine the peak luminosity as
\begin{gather}
    \mathcal{L}_{\mathrm{peak}} = \smash{\displaystyle\max_{t}}
    \, \mathcal{L} (t) \, ,
\label{eq:peak_lum}
\end{gather}
where we determine the peak value by fitting a quadratic function to 5 adjacent
samples of $\mathcal{L}(t)$, consisting of the largest sample and two neighbors
on either side. Before applying our fitting method, we first take a logarithm
of the peak luminosity and model $\log{(\mathcal{L}_{\mathrm{peak}})}$. We find
that this leads to more accurate fits than directly modeling
$\mathcal{L}_{\mathrm{peak}}$. When the model is evaluated, we can easily get
the predicted peak luminosity by taking the exponential of the fit output.

For the aligned-spin model \texttt{NRSur3dq8Remnant}, we include the
$\ell\leq4$ and (5,5) modes but not the (4,1) or (4,0) modes in
Eq.~(\ref{eq:luminosity}). We include the $m>0$ modes twice to account for the
$m<0$ modes, which are given by $\h_{\ell, -m} = (-1)^{\ell} \, \hlm^*$ due to
the symmetries of aligned-spin systems.  The included modes are the same as
those used for the surrogate model of Ref.~\cite{Varma:2018mmi}. The reason for
excluding the (4,1), (4,0), $(\ell=5, m<5)$ and $\ell>5$ modes is two fold: (i)
These modes have very small amplitudes and do not contribute significantly to
the sum in Eq.~(\ref{eq:luminosity}). (ii) The small amplitude of some of these
modes (particularly (4,1) and (4,0)) can behave poorly when
extrapolated~\cite{Boyle:2009vi}. We expect that this will be resolved in the
future with Cauchy characteristic extraction~\cite{Barkett:2019uae,
Moxon:2020gha}.

For the precessing model \texttt{NRSur7dq4Remnant}, we use all
$\ell\leq5$ modes. Due to the orbital precession, even modes like (4,1), (4,0)
and $(\ell=5, m<5)$ can have significant amplitude due to mode mixing (see for
e.g.  Ref.~\cite{Varma:2019csw}). Therefore, these modes behave reasonably well
when extrapolated. Note that the $m<0$ modes are directly included when doing
the sum in Eq.~(\ref{eq:luminosity}) as the aforementioned symmetry for $m<0$
does not hold for precessing systems.

\subsection{Gaussian process regression}
\label{sec:gpr_fits}
We construct fits in this work using GPR~\cite{Rasmussen_Williams_GPRbook} as
implemented in {\it scikit-learn}~\cite{scikit}. We closely follow the
procedure outlined in the supplement of Ref.~\cite{Varma:2018aht}, which we
describe briefly in the following.

We start with a training set of $n$ observations, $\ts = \left\{\bLambda^i,
f(\bLambda^i)) | i = 1,\dots,n\right\}$, where each $\bLambda^i$ denotes an
input vector of dimension $D$ and $f(\bLambda^i)$ is the corresponding scalar
output.  In our case, $\bLambda$ is given by Eq.(\ref{eq:lambda_precessing})
and Eq.(\ref{eq:lambda_aligned}) respectively, for the precessing and
aligned-spin models, and $f(\bLambda)=\log{(\mathcal{L}_{\mathrm{peak}})}$. Our
goal is to use $\ts$ to make predictions for the underlying $f(\bLambda)$ at
any point $\bLambda_*$ that is not in $\ts$.

A Gaussian process (GP) can be thought of as a probability distribution of
functions. More formally, a GP is a collection of random variables, any finite
number of which have a joint Gaussian
distribution~\cite{Rasmussen_Williams_GPRbook}.  A GP is completely specified
by its mean function $m(\bLambda)$ and covariance function
$k(\bLambda,\bLambda^\prime)$, i.e.  $f(\bLambda) \sim
\mathcal{GP}(m(\bLambda), k(\bLambda, \bLambda^\prime))$. Consider a prediction
set of $n_*$ test inputs and their corresponding outputs (which are unknown):
$\ps = \left\{(\bLambda^i_*, f(\bLambda^i_*)) | i = 1,\dots,n_*\right\}$.  By
the definition of a GP, outputs of $\ts$ and $\ps$ (respectively ${\bf
    f}\!=\!\{f(\bLambda^i)\}$, ${\bf f_*}\!=\!\{f(\bLambda^i_*)\}$) are related
    by a joint Gaussian distribution:
\begin{gather}
\begin{bmatrix} {\bf{f}} \\ {\bf{f_*}} \end{bmatrix} =
    \mathcal{N}\left({\bf 0},
        \begin{bmatrix}
            K_{\bLambda \bLambda} & K_{\bLambda \bLambda_*} \\
            K_{\bLambda_* \bLambda} & K_{\bLambda_* \bLambda_*}
        \end{bmatrix}
    \right),
\label{Eq:GPR_prior}
\end{gather}
where $K_{\bLambda \bLambda_*}$ denotes the $n \into n_*$ matrix of the
covariance $k(\bLambda,\bLambda_*)$ evaluated at all pairs of training and
prediction points, and similarly for the other $K$ matrices.

Eq.~(\ref{Eq:GPR_prior}) provides the Bayesian prior distribution for ${\bf
f_*}$. The posterior distribution is obtained by restricting this joint prior
to contain only those functions which agree with the observed data
points~\cite{Rasmussen_Williams_GPRbook}, i.e.
\begin{align}
p({\bf f_*}|\ts) = \mathcal{N} \Bigg(&
K_{\bLambda_* \bLambda} ~ K_{\bLambda \bLambda}^{-1} ~ {\bf f} \,, \nonumber \\
& \, K_{\bLambda_* \bLambda_*} - K_{\bLambda_* \bLambda}
~ K_{\bLambda \bLambda}^{-1} ~ K_{\bLambda \bLambda_*}
\Bigg).
\label{Eq:GPR_posterior}
\end{align}
The mean of this posterior provides an estimator for $f(\bLambda)$ at
$\bLambda_*$, while its width is the prediction error.

Finally, one needs to specify the covariance (or kernel) function
$k(\bLambda,\bLambda^\prime)$. Following Ref.~\cite{Varma:2018aht}, we
implement the following kernel
\begin{gather}
k(\bLambda, \bLambda^\prime) =\sigma_{k}^2 \exp{\left[\! -\sum_{j=1}^{D}
\left(\frac{\bLambda^j \!-\! \bLambda^{\prime j}}{\sqrt{2} \, \sigma_j}\right)^2
\!\right]} + \sigma_n^2 \, \delta_{\bLambda\bLambda'}\,,
\label{Eq:GPR_kernel}
\end{gather}
where $\delta_{\bLambda\bLambda^{\prime}}$ is the Kronecker delta. In words, we
use a product between a squared exponential kernel (parametrized by $\sigma_j$)
and a constant kernel (parametrized by $\sigma_k^2$), to which we add a white
kernel (parametrized by $\sigma_n^2$) to account for additional noise in the
training data \cite{Rasmussen_Williams_GPRbook,scikit}.

GPR fit construction involves determining the $D\!+\!2$ hyperparameters
($\sigma_k$, $\sigma_n$ and $\sigma_j$) which maximize the marginal likelihood
of the training data under the GP prior~\cite{Rasmussen_Williams_GPRbook}.
Local maxima are avoided by repeating the optimization with 10 different
initial guesses, obtained by sampling uniformly in log in the hyperparameter
space described below.

Before constructing the GPR fit, we pre-process the training data as follows.
We first subtract a linear fit and the mean of the resulting values. The data
are then normalized by dividing by the standard deviation of the resulting
values. The inverse of these transformations is applied at the time of the fit
evaluation. The reasoning behind the pre-processing is two-fold: (1) The
de-meaning and normalization allows us to apply the same ranges (described
below) for the GPR hyperparameters for a wide range of models. For instance, we
used the same settings to model the remnant BH properties in
Refs.~\cite{Varma:2018aht, Varma:2019csw}. (2) The data are simpler to model
after removing the linear component, leading to more accurate fits.

For each dimension of $\bLambda$, we define $\Delta \bLambda^j$ to be the range
of the values of $\bLambda^j$ in $\ts$ and consider $\sigma_j \in [0.01 \times
\Delta \bLambda^j, 10\times \Delta \bLambda^j]$. Larger length scales are
unlikely to be relevant and smaller length scales are unlikely to be
resolvable. The remaining hyperparameters are sampled in $\sigma_k^2\in
[10^{-2}, 10^2]$ and  $\sigma_n^2 \in [10^{-7}, 10^{-2}]$. These choices are
meant to be conservative and are based on prior exploration of the typical
magnitude and noise level in our training data.

\subsection{Precessing model, \texttt{NRSur7dq4Remnant}}
\label{sec:precessing_model}

For precessing systems the parameter space is 7-dimensional comprising of the
mass $q$, and two spin 3-vectors $\bm{\chi}_1$ and $\bm{\chi}_{2}$. Here
$q=m_1/m_2$ is the mass ratio with $m_1 \geq m_2$, and $\bm{\chi}_{1}$
($\bm{\chi}_{2}$) is the dimensionless spin vector of the heavier (lighter) BH.
The total mass ($M=m_1+m_2$) scales out of the problem and does not constitute
an additional parameter for modeling. We use the 1528 NR waveforms used for the
surrogate models of Ref.~\cite{Varma:2019csw}, which cover the parameter space
$q\leq4$, $\chi_1,\chi_2\leq0.8$, where $\chi_1$ ($\chi_2$) is the magnitude of
$\bchi_1$ ($\bchi_2$).

Following Refs.~\cite{Varma:2019csw, Varma:2018aht}, we parametrize the
precessing fit using the coorbital frame spins $\bchiCoorb_{1,2}$ at
$\tmHundred$ before the peak of the total waveform amplitude (as defined in
Eq.~5 of Ref.~\cite{Varma:2019csw}). The coorbital frame is a time-dependent
non-inertial frame in which the $z-$axis is along the instantaneous
$\hat{\bm{L}}$ direction, and $x-$axis is along the instantaneous
line-of-separation between the BHs, with the heavier BH on the positive
$x-$axis.\footnote{Here the BH positions are defined using the waveform at
future null infinity and do not necessarily correspond to the (gauge-dependent)
coordinate BH positions in the NR simulation. See Ref.~\cite{Varma:2019csw} for
more details.} The \texttt{NRSur7dq4Remnant} fit is parametrized as follows.
\begin{equation}
\bLambda = [\log(q), \chiCoorb_{1x}, \chiCoorb_{1y}, \chihatCoorb,
 \chiCoorb_{2x}, \chiCoorb_{2y}, \chiCoorb_{a}],
\label{eq:lambda_precessing}
\end{equation}
where $\chihatCoorb$ is the spin parameter entering the GW phase at leading
order \cite{Khan:2015jqa, Ajith:2011ec, CutlerFlanagan1994, Poisson:1995ef} in
the PN expansion
\begin{gather}
\label{eq:chihat_coorb}
\chihatCoorb = \frac{\chieffCoorb - 38\eta(\chiCoorb_{1z}
     + \chiCoorb_{2z})/113} {1-{76\eta}/{113}} \, , \\
\chieffCoorb = \frac{q~\chiCoorb_{1z} + \chiCoorb_{2z}}{1+q} , \\
\eta = \frac{q}{(1+q)^2} \, ,
\end{gather}
and $\chiCoorb_a$ is the ``anti-symmetric spin'',
\begin{equation}
    \chiCoorb_a = \tfrac{1}{2}(\chiCoorb_{1z} - \chiCoorb_{2z})\,.
\label{eq:chia_coorb}
\end{equation}
We empirically found this parameterization to perform more accurately than the
more intuitive choice $\tilde{\bLambda}=[q, \chiCoorb_{1x}, \chiCoorb_{1y},
\chiCoorb_{1z}, \chiCoorb_{2x}, \chiCoorb_{2y}, \chiCoorb_{2z}]$ used in
Ref.~\cite{Blackman:2017pcm}.

\subsection{Aligned-spin model, \texttt{NRSur3dq8Remnant}}
\label{sec:aligned_model}

\texttt{NRSur7dq4Remnant} is restricted to $q\leq4$ due to a lack of sufficient
precessing simulations at higher mass ratios~\cite{Boyle:2019kee}. NR
simulations become increasingly expensive as one approaches higher mass ratios
and/or spin magnitudes. However, the SXS Catalog has good coverage for
aligned-spin BBHs up to $q\leq8$~\cite{Varma:2018mmi, Boyle:2019kee}. We make
use of the 104 NR waveforms used for the surrogate model of
Ref.~\cite{Varma:2018mmi}, which cover the parameter space $q\leq8$,
$|\chi_{1z}|,|\chi_{2z}|\leq0.8$.

Note that the spins in aligned-spin BBHs are restricted to the $\hat{\bm{L}}$
direction, this reduces the parameter space to 3-dimensions.  Following
Refs.~\cite{Varma:2018aht, Varma:2018mmi}, we parametrize the
\texttt{NRSur3dq8Remnant} fit as follows.
\begin{equation}
    \bLambda = [\log(q), \chihatCoorb, \chiCoorb_{a}],
\label{eq:lambda_aligned}
\end{equation}
where, we use Eq.~(\ref{eq:chihat_coorb}) and (\ref{eq:chia_coorb}), but
keeping in mind that spins in the coorbital-frame are the same as those in the
inertial frame for aligned-spin systems.

\section{Modeling errors}
\label{sec:errors}

We evaluate the accuracy of our new surrogate models by comparing against the
the NR simulations used in this work. To avoid underestimating the errors, we
perform a 20-fold cross-validation study to compute “out-of-sample” errors as
follows.  We first randomly divide the training simulations into 20 groups of
roughly the same size. For each group, we build a trial surrogate using the
remaining training simulations and test against the simulations in that group,
which may include points on the boundary of the training set.

For comparison, we also compute the errors for existing peak luminosity fitting
formulae~\cite{Keitel:2016krm, Healy:2016lce, JimenezForteza2016} against the
NR simulations. We refer to the fit of Ref.~\cite{Keitel:2016krm} as
UIB~\footnote{After the research group.}, the fit of Ref.~\cite{Healy:2016lce}
as HL~\footnote{For the authors Healy+Lousto.}, and the fit of
Ref.~\cite{JimenezForteza2016} as FK~\footnote{For the lead authors
Forteza+Keitel.}. Note that these fits are not trained on precessing
simulations. As the spins evolve for precessing systems, there is an ambiguity
about at what time these fits should be evaluated. We follow the procedure
outlined in Ref.~\cite{mcdaniel2016} and used in LIGO/Virgo analyses
(e.g.~\cite{Abbott:2020tfl}): NR spins are evolved from relaxation to the
Schwarzschild innermost stable circular orbit (ISCO) using post-Newtonian (PN)
theory. The spins at ISCO, projected along $\hat{\bm{L}}$, are used to evaluate
the aligned-spin peak luminosity fitting formulae.

\subsection{Errors for the precessing model}

\begin{figure}[thb]
\includegraphics[width=0.9\columnwidth]{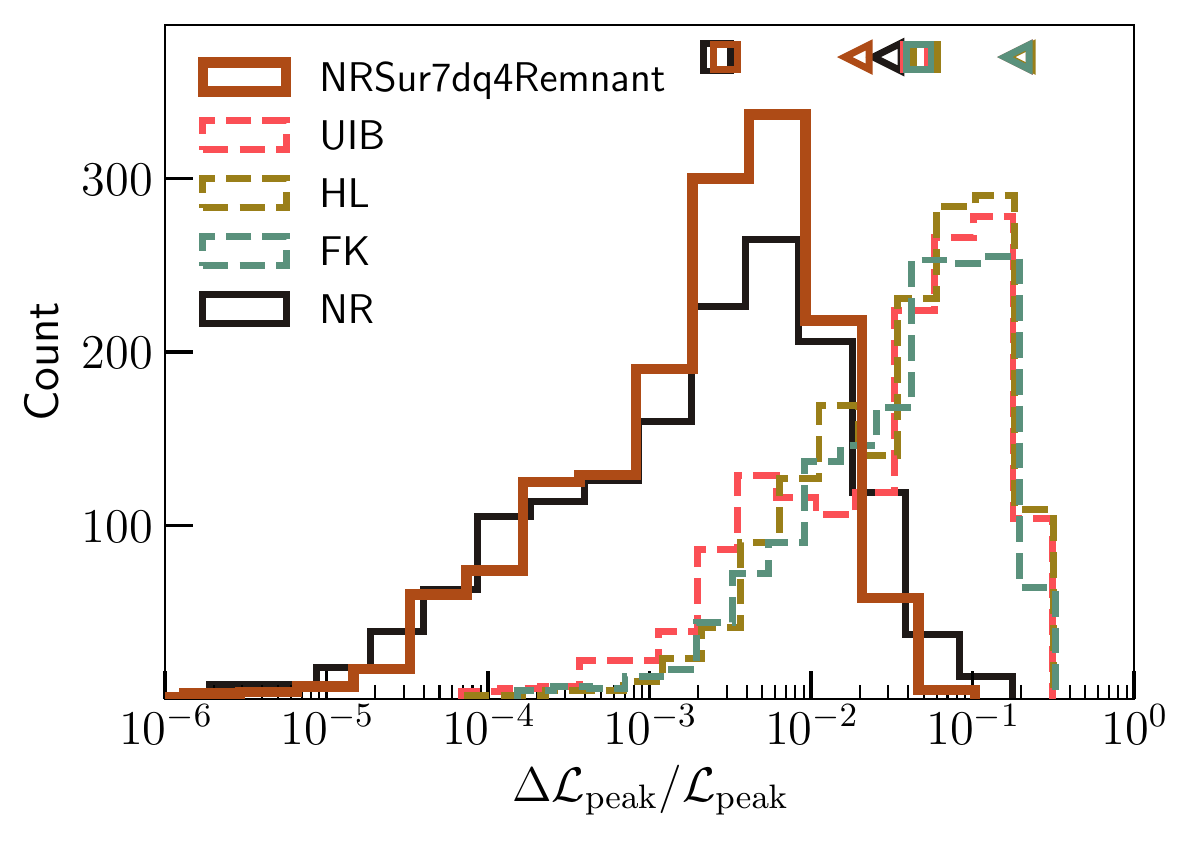}
\label{fig:errors_precessing}
\caption{Fractional errors (out-of-sample) in predicting the peak luminosity
    for the precessing model \texttt{NRSur7dq4Remnant} when compared against
    precessing NR simulations.  When evaluating \texttt{NRSur7dq4Remnant}, we
    use the NR spins at $\tmHundred$, where the model was trained. Also shown
    are the NR resolution errors and errors for different existing fitting
    formulae. The square (triangle) markers at the top indicate the median
    (95th percentile) values. \texttt{NRSur7dq4Remnant} is more accurate than
    the existing formulae by about an order of magnitude.
}
\end{figure}

\begin{figure}[thb]
\includegraphics[width=0.8\columnwidth]{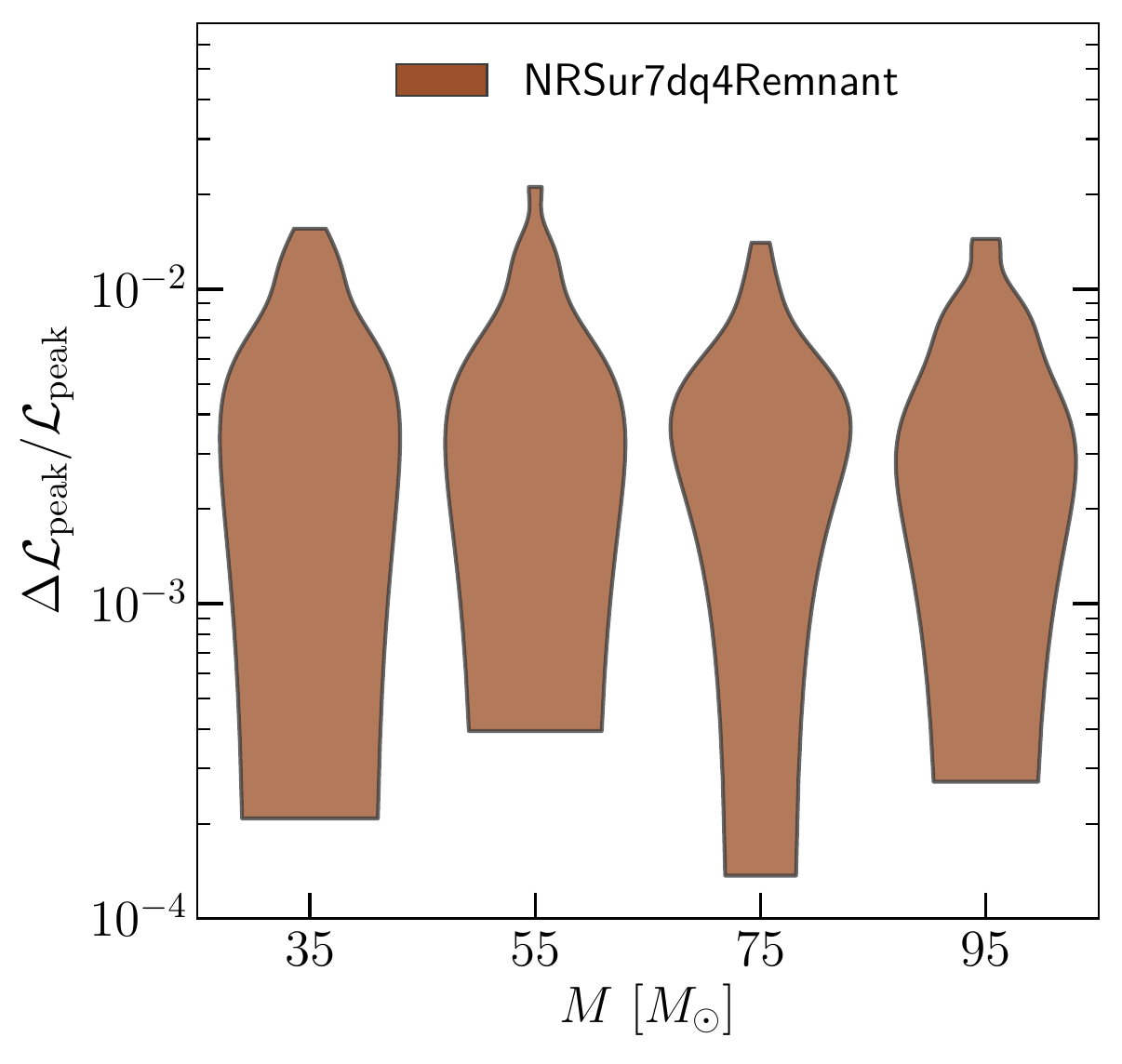}
\label{fig:errors_precessing_spin_evolve}
\caption{Fractional errors for \texttt{NRSur7dq4Remnant} in predicting the peak
    luminosity when spins are specified at a reference frequency of
    $f_{\mathrm{ref}}=20$ Hz. For four different total masses, we compute the
    errors against 23 long NR simulations that were not used to train the
    model. For each mass, the errors are shown as a smoothed vertical histogram
    (or a violin). The histograms are normalized so that all violins have equal
    width.
}
\end{figure}

We demonstrate the accuracy of the \texttt{NRSur7dq4Remnant} model by comparing
against the 1528 precessing NR simulations described in
Sec.~\ref{sec:precessing_model}. We perform a 20-fold cross-validation study
where we leave out $\sim75$ simulations in each trial for testing.
Figure.~\ref{fig:errors_precessing} shows the errors for
\texttt{NRSur7dq4Remnant} when using the NR spins at $\tmHundred$ as the input.
As the model was trained at this time, these errors represent the errors in the
GPR fitting procedure. The 95th percentile fractional error in predicting the
peak luminosity is $\roughly0.02$. We also show the errors for existing fitting
formulae, and the NR resolution error, estimated by comparing the two highest
resolution simulations. Our errors are at the same level as the estimated NR
error, and about an order of magnitude smaller than that of existing fitting
formulae.

In practice, one might want to specify the input spins at arbitrary times. For
example, in LIGO-Virgo analyses (e.g.~\cite{LIGOScientific:2018mvr}) the spins
are measured at a fixed reference frequency. Following
Refs.~\cite{Varma:2018aht, Varma:2019csw} this is handled by evolving the input
spins from the reference frequency to $\tmHundred$ using a combination of PN in
the early inspiral and NRSur7dq4~\cite{Varma:2019csw} spin evolution in the
late inspiral.  Figure~\ref{fig:errors_precessing_spin_evolve} shows the errors
in \texttt{NRSur7dq4Remnant} when the spins are specified at a reference
orbital frequency $f_{\mathrm{ref}}=20$ Hz. These errors are computed by
comparing against 23 long NR ($3\times10^4M$ to $10^5M$ in length)
simulations~\cite{SXSCatalog2018, Varma:2019csw} with mass ratios $q\leq4$ and
generically oriented spins with magnitudes $\chi_1, \chi_2 \sim 0.5$. Note that
none of these simulations were used to train the surrogates. Comparing with
Fig.~\ref{fig:errors_precessing}, even with spin evolution, our errors are
about an order of magnitude lower than that of existing fits.

\subsection{Errors for the aligned-spin model}

\begin{figure}[thb]
\includegraphics[width=0.9\columnwidth]{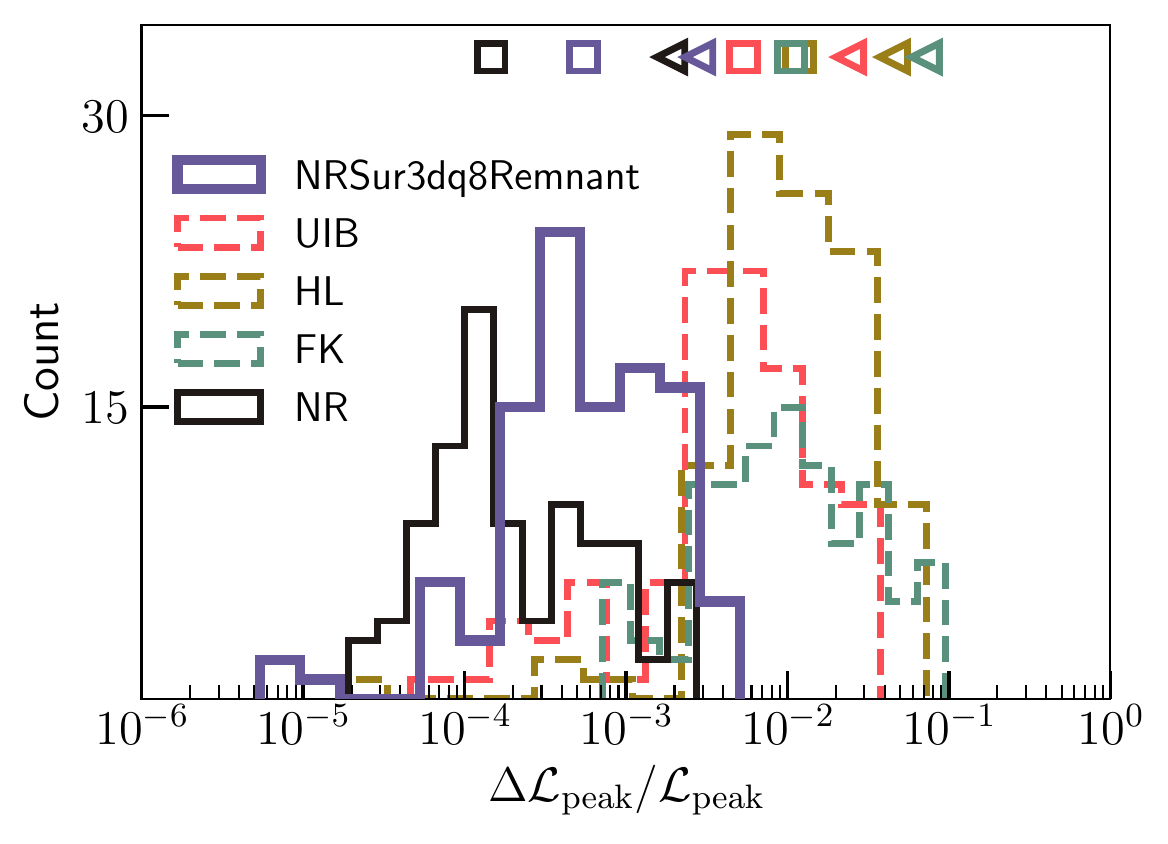}
\label{fig:errors_aligned}
\caption{Fractional errors (out-of-sample) in predicting the peak luminosity
    for the aligned-spin model \texttt{NRSur3dq8Remnant} when compared against
    aligned-spin NR simulations. Also shown are the NR resolution errors and
    errors for different existing fitting formulae. The square (triangle)
    markers at the top indicate the median (95th percentile) values.
    \texttt{NRSur3dq8Remnant} is more accurate than the existing formulae by at
    least an order of magnitude.
}
\end{figure}

We demonstrate the accuracy of the \texttt{NRSur3dq8Remnant} model by comparing
against the 104 aligned-spin NR simulations described in
Sec.~\ref{sec:aligned_model}. We perform a 20-fold cross-validation study where
we leave out $\sim5$ simulations in each trial for testing. These errors are
shown in Fig.~\ref{fig:errors_aligned}. The 95th percentile fractional error in
predicting the peak luminosity is $\roughly0.002$.
Fig.~\ref{fig:errors_aligned} also shows the errors for the existing fitting
formulae and the estimated NR errors. \texttt{NRSur3dq8Remnant} is comparable
to NR and more accurate than existing fits by at least an order of magnitude.

We note that the estimated NR errors for aligned-spin BBHs
(Fig.~\ref{fig:errors_aligned}) are significantly smaller than that for
precessing BBHs (Fig.~\ref{fig:errors_precessing}). The reason for this is not
clear, but this places a limit on how accurate the surrogate models can be.
This is reflected in the higher errors for \texttt{NRSur7dq4Remnant} compared
to \texttt{NRSur3dq8Remnant}. More accurate precessing NR simulations may be
necessary to further improve the precessing model.

\begin{figure}[thb]
\includegraphics[width=0.9\columnwidth]{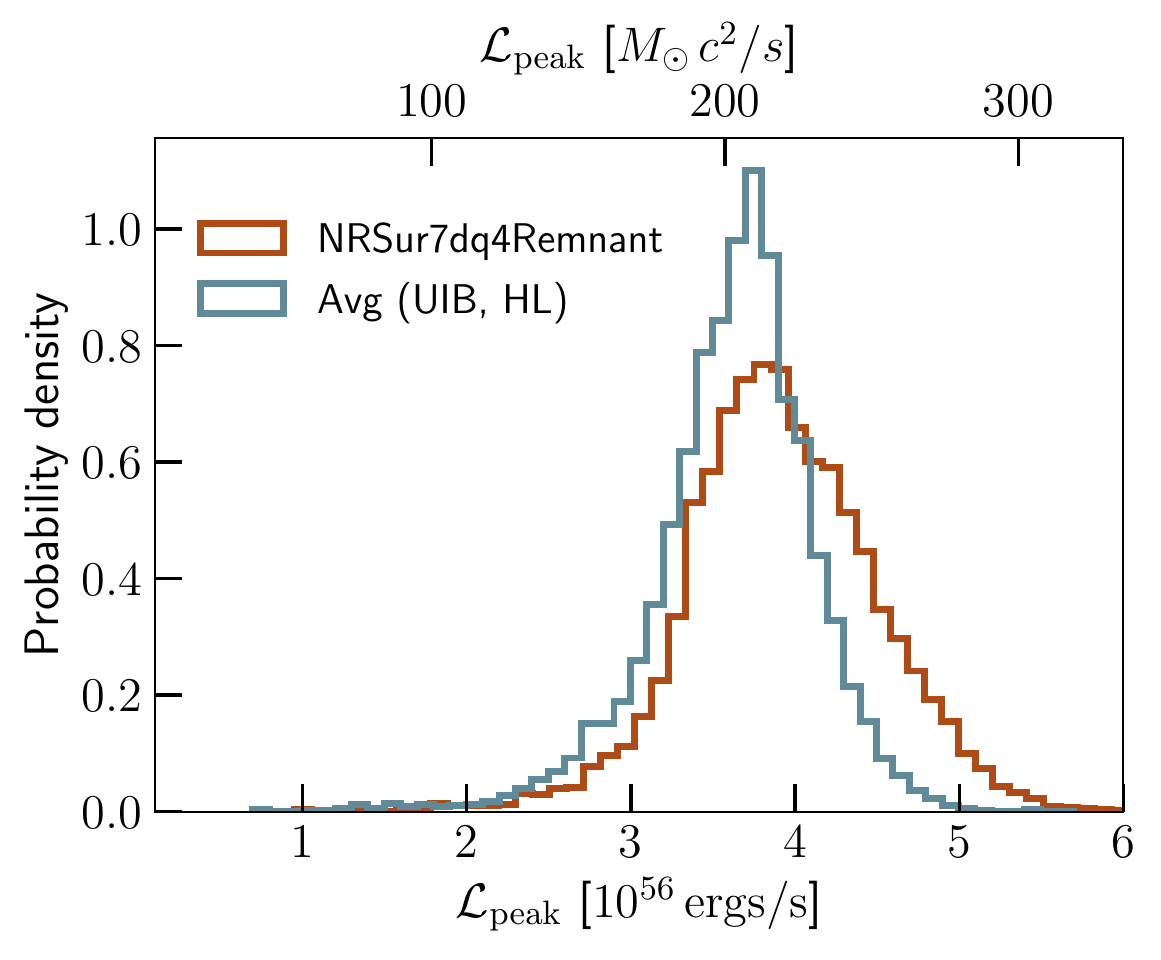}
\label{fig:GW190521_peaklum}
\caption{Posterior distribution for the peak luminosity of GW190521, obtained
    using the \texttt{NRSur7dq4Remnant} model as well as the average of the
    UIB~\cite{Keitel:2016krm} and HL~\cite{Healy:2016lce} fitting formulae.
    While the two posteriors are consistent with each other,
    \texttt{NRSur7dq4Remnant} suggests a slightly higher value for the peak
    luminosity.
}
\end{figure}

\section{Peak luminosity of GW190521}
\label{sec:GW190521_peak_lum}

As a first application of our models, we compute the peak luminosity of
GW190521~\cite{Abbott:2020tfl} using \texttt{NRSur7dq4Remnant}. We apply the
\texttt{NRSur7dq4Remnant} model to the posteriors samples for the component
masses and spins, obtained using the preferred \texttt{NRSur7dq4} model in
Ref.~\cite{Abbott:2020mjq}, and made publicly
available~\cite{GW_open_science_center} by the LIGO-Virgo Collaboration. This
peak luminosity posterior is shown in Fig.~\ref{fig:GW190521_peaklum}. We
compare this with the peak luminosity posterior obtained in
Ref.~\cite{Abbott:2020mjq} using the average of the UIB~\cite{Keitel:2016krm}
and HL~\cite{Healy:2016lce} fitting formulae applied to the same
\texttt{NRSur7dq4} posterior samples. While the two posteriors are consistent
with each other, \texttt{NRSur7dq4Remnant} shows support for slightly higher
values of peak luminosity. This level of agreement is expected, as GW190521 had
a relatively weak signal-to-noise ratio of $\roughly
14.7$~\cite{Abbott:2020tfl}.  As GW detectors become more sensitive in the
coming years, we can expect to see stronger signals for which systematic biases
in peak luminosity models will become important.

\section{Conclusion}
\label{sec:conclusion}
We present GPR based NR surrogate models for peak luminosity of BBH mergers.
The first model, \texttt{NRSur7dq4Remnant}, is trained on 1528 precessing
systems with mass ratios $q\leq4$ and spin magnitudes $\chi_{1},\chi_{2}
\leq0.8$. The second model, \texttt{NRSur3dq8Remnant}, is trained on 104
aligned-spin systems with mass ratios $q\leq8$ and spins
$|\chi_{1z}|,|\chi_{2z}|\leq0.8$. Both models are comparable to the NR
simulations in accuracy, and outperform existing fitting formulate by an order
of magnitude or more. The models are made publicly available through the
Python module \texttt{surfinBH}~\cite{surfinBH} and can be used to estimate the
peak luminosity of GW signals. We use \texttt{NRSur7dq4Remnant} to infer the
peak luminosity of the GW event GW190521, and find the results to be consistent
with previous predictions.

As our GW detectors improve, we will need models that capture the full physics
of BBH mergers. \texttt{NRSur7dq4Remnant} is the first peak luminosity model
trained on precessing NR simulations. Models such as this will become
necessary to accurately infer the peak luminosity as we approach the era of
high-precision GW astronomy.

\begin{acknowledgments}
We thank Scott Field, Leo Stein and Carl-Johan Haster for useful comments.
This research made use of data, software and/or web tools obtained from the
Gravitational Wave Open Science Center~\cite{GW_open_science_center}, a service
of the LIGO Laboratory, the LIGO Scientific Collaboration and the Virgo
Collaboration.
A.T.\ gratefully acknowledges the support of the United States National Science
Foundation (NSF) for the construction and operation of the LIGO Laboratory and
Advanced LIGO, the LIGO Laboratory NSF Research Experience for Undergraduates
(REU) program, and the Carl A. Rouse Family.
V.V.\ is generously supported by a Klarman Fellowship at Cornell, the
Sherman Fairchild Foundation, and NSF grants PHY–170212 and PHY–1708213 at
Caltech.
\end{acknowledgments}

\section*{References}
\bibliography{References}

\end{document}